\documentclass[]{aa} 
\usepackage{graphicx}
\usepackage{txfonts}
\usepackage{natbib}
\def\eg{{\it e.g.~}}

\def\ie{{\it i.e.~}}
\def\km2{\ensuremath{\mathrm{km}^{2}}}
\def\V{\ensuremath{V}~}
\def\R{\ensuremath{R}~}
\def\I{\ensuremath{I}~}
\def\VRI{\ensuremath{VRI}~}
\def\SAf{\ensuremath{\Sigma Af}}
\def\Af{\ensuremath{Af}}
\def\Afrho{\ensuremath{Af\rho}}
\def\rh{\ensuremath{\mathrm{r_h}}}
\def\Ll{\ensuremath{L_\mathrm{l}}}
\def\Ls{\ensuremath{L_\mathrm{s}}}
\def\Nl{\ensuremath{N_\mathrm{l}}}
\def\Ns{\ensuremath{N_\mathrm{s}}}
\def\Nc{\ensuremath{N_\mathrm{c}}}
\def\N0{\ensuremath{N_\mathrm{0}}}
\def\Nd{\ensuremath{N\mathrm{(d)}}}

\begin{document}

\title{Evolution of the Dust Coma in Comet 67P/Churyumov-Gerasimenko Before 2009 Perihelion}

\author{G. P. Tozzi \inst{1} \and P. Patriarchi \inst{1} \and H. Boehnhardt \inst{2} and J.-B. Vincent \inst{2} \and J. Licandro \inst{3,4} \and L. Kolokolova \inst{5} \and R. Schulz \inst{6} and J. St\"uwe  \inst{6} \fnmsep \thanks{Based on observations collected at the European Organization for Astronomical Research in the Southern Hemisphere, Chile 
(programs 381.C-0123 \& 082.C-0740) and Telescopio Nazionale Galileo (TNG) of INAF (program TAC\_35)}}
\institute{INAF--Osservatorio Astrofisico di Arcetri, Largo E. Fermi 5, I-50125, Firenze (I)
\and Max- Planck Institut f\"ur Sonnensystemforschung, 37191 Katlenburg-Lindau (D) \and Instituto de Astrof\'{\i}sica de Canarias, V\'{\i}a L\'actea s/n, 38200 La Laguna, Tenerife (E). 
              \and
              Departamento de Astrof\'{\i}sica, Universidad de La Laguna, E-38205 La Laguna, Tenerife, (E)  \and University of Maryland, College Park, MD 20742-2421 (USA) \and ESA Research and Scientific Support Department, ESTEC, 2200 AG Noordwijk (NL)}

\date{Received xxx; accepted xxx}

\abstract
{Comet 67P/Churyumov-Gerasimenko is the main target of ESA's Rosetta mission and  will be encountered in May 2014. As the spacecraft shall be in orbit the comet nucleus before and after release of the lander {\it Philae}, it is
necessary necessary to know the conditions in the coma}
{Study the dust environment, including the dust production rate and its variations along its preperihelion orbit.}
{The comet was observed during its approach to the Sun on four epochs between early-June 2008 and mid-January 2009, over a large range of heliocentric distances that will be covered by the mission in 2014.} 
{An anomalous enhancement of the coma dust density was measured towards the comet nucleus. The scalelength of this enhancement increased with decreasing heliocentric distance of the comet. This is interpreted as a result of an unusually slow expansion of the dust coma. Assuming a spherical symmetric coma, the average amount of dust as well as its ejection velocity have been derived. The latter increases exponentially with decreasing heliocentric distance (\rh), ranging from about 1 m/s at 3 AU to about 25-35 m/s at 1.4 AU. Based on these results we describe the dust environment at those nucleocentric distances at which the spacecraft will presumably be in orbit.}
{}

\maketitle

\section{Introduction}

 In May 2014 ESA's {\it Rosetta} spacecraft, an orbiter and lander mission to a comet, will encounter its 
target, comet 67P/Churyumov-Gerasimenko (hereafter 67P), at a heliocentric distance \rh\ of about 4 AU. It will go in  orbit around the nucleus in September 2014, when the comet is at \rh = 3.4 AU and start global observations and mapping of the nucleus \citep{Glassmeier2007} . At \rh $\approx$ 3 AU, the lander {\it Philae} 
will land on the  surface and perform the first-ever in-situ analysis of comet nucleus material. The orbiter will be  monitoring the evolution of the nucleus and the coma along the comet's pre- and post-perihelion orbit 
for more than one year while approaching the nucleus as close as a few km above the surface. 

When approaching the Sun, at distances from about 5 to 2 AU, cometary nuclei usually switch on the activity and 
produce a gas and dust coma characterized by a changing environment due to secular and short-term effects, e.g., solar heating and rotation of the nucleus. Consequently, an increase of the 
production rate and expansion speed of the dust is expected as a result of the increased heating of sublimating ices from the nucleus. 
Unfortunately not much is published on the coma status and the dust environment of comets at large distances, because comets are very faint when far from the Sun, and require observations with large telescopes for which observing time is  difficult to receive. This 
applies particularly to the short-period comets of the Jupiter-family, of which the {\it Rosetta} target comet
67P is a member. At large \rh\ the \Afrho\ index \citep{AHearn1984} is a questionable measure for the dust production rate because the conditions for radially constant \Afrho\ profiles 
(homogeneous, isotropic and constant outflow of the dust) are not fulfilled. It can be suspected that the expansion velocities
of the dust in the coma are low, resulting in very long travel times (weeks to months) for the grains through the coma, particularly for the larger grains. This may lead to a pile-up of grains in the 
coma, reflecting the changing dust production of the nucleus over a range of solar distances. 

Apart from the scientific interest in obtaining a better knowledge of the dust coma 
environment and evolution of 67P along the preperihelion orbit, this knowledge is required for the safety and the operations planning of the {\it Rosetta} mission. 
After the {\it Rosetta} launch delay in 2003, 67P was selected as the new target of the mission replacing the original 
target 46P/Wirtanen which was much better characterized by Earth-based observations. At the time of its 
selection 67P had just passed its perihelion. Thus, most observations existing to date were obtained during the 
subsequent post-perihelion phase \citep{Schulz2004, Weiler2004}, whereas only a few observations exist of 
the pre-perihelion phase \citep{Schleicher2006, Lamy2006, Lamy2007, Lamy2008}, except from those obtained by amateur astronomers 
(cara.uai.it, www.aerith.net), \citep{Kidger2003, Kidger2004}. 

The 2009 return of the comet 
was the first and last opportunity to measure the pre-perihelion dust environment of 67P before
{\it Rosetta} will have arrived at the comet.  

Here, we report the results of our dust coma analysis of 67P using observations obtained during the pre-perihelion 
phase 2008-2009 between \rh $ \approx$ 3 to 1.4 AU. Our observations are 
complementary to those reported by \cite{Tubiana2008, Tubiana2010}, obtained at larger heliocentric distances and those by \cite{Lara2010}, obtained during the post-perihelion phase.

\section{Observations}

Comet 67P was observed at four epochs (see Table \ref{tablog}) during
the inbound phase before it reached its perihelion in Feb 2009. During
the first three epochs the observations were obtained at the 8.2m Very
Large Telescope (VLT) of the European Southern Observatory in Chile using
the FORS2 focal reducer instrument (see
www.eso.org/sci/facilities/paranal/instruments). The last run, when
the comet was in the northern hemisphere, was performed with the
Telescopio Nazionale Galileo TNG at La Palma using the
camera-spectrograph DOLORES (see www.tng.iac.es/instruments/).  During each
run a series of images was acquired using broadband
filters (\VRI bands) and sometimes narrower band filters,  taken a few
days apart in order to check the comet short term
variability. Differential tracking of the telescope at the cometary
velocity was applied for all 67P images.  Since the comet was in front
of star-rich regions and since it had a relatively fast
proper motion, each run consisted of several (5-9) exposures in order
to reduce the star contribution by median averaging of the
images during data processing. At the VLT the observations were
performed in service mode and the flux calibration relies on the zero
point estimations (Zp) provided by the observatory for each
photometric night. The observations at the TNG were taken in visitor mode and
a series of photometric standards were obtained for calibration purposes.

\begin{table}
\caption{\label{tablog} Summary log for the 
observations of comet 67P.  The table lists the date of observations (YYMMDD), the heliocentric \rh\ and
geocentric distance $\Delta$ of the comet in AU, the phase angle ($\alpha$), and the position angle
of the extended Sun--Comet radius vector (PA$_{tail}$) and of the velocity vector (PA$_{Vel.}$). 
The last column gives the multiplicative factor to apply to transform
the measurements to a phase equal to 0\degr\ \citep{Schleicher2010}}

\begin{center}
\begin{tabular}{cccccccc}
\hline
\hline
Date & r$_h$ & $\Delta$ & $\alpha$ & PA$_{tail}$ &
PA$_{Vel.} $ & Coeff. \\ 
YYMMDD & AU &  AU &  $\degr$ & $\degr$ &
$\degr$ &  \\ 
\hline
080601 & 2.98 & 2.50 & 18.83 & 252.9 & 250.3& 1.89\\ 
080604 & 2.96 & 2.44 & 18.67 & 253.3 & 250.2& 1.89\\ 
080605 & 2.95 & 2.42 & 18.60 & 253.4 & 250.2& 1.88\\ 
080904 & 2.30 & 1.40 & 14.30 & ~55.7 & 260.2& 1.67\\ 
080906 & 2.28 & 1.39 & 15.23 & ~57.4 & 260.3& 1.71\\ 
080908 & 2.27 & 1.39 & 16.14 & ~59.0 & 260.6& 1.76\\ 
081022 & 1.93 & 1.51 & 30.63 & ~73.0 & 259.0& 2.44\\
081026 & 1.90 & 1.53 & 31.37 & ~73.3 & 258.1& 2.47\\
090113 & 1.36 & 1.67 & 35.98 & ~68.7 & 239.1& 2.64\\
\hline
\end{tabular}
\end{center}
\end{table}

\section{Data reduction}

All images were processed using standard CCD reduction procedures
(BIAS \& Flat Field correction).  Then a first-order constant sky
level, measured in image regions away from the comet's photocenter, was
subtracted.  After combining all filter frames of a single night
(aligned to the comet and median averaged), possible sky residuals
were checked in the resulting frame and subtracted using the \SAf~
function (see below). Finally, the images of each night were
calibrated in \Af, \ie the albedo multiplied by the filling factor of
the dust in the coma \citep{AHearn1984, Tozzi2007}. In order to allow
comparison of the results obtained at different phase angles, the
calibrated images were re-calibrated for the phase equal to
0\degr\ using the multiplicative phase correction coefficient (last
column of Table \ref{tablog}) given by \cite{Schleicher1998} and
recently refined in \citep{Schleicher2010}. The observation series
taken over a few days apart were searched for the presence of possible
short-term variability. Since day-to-day variability was not found,
all images obtained with the same filter during each epoch were median
averaged, thus further reducing the contribution of the background
stars and increasing the signal-to-noise ratio of the cometary coma
flux.

These final images in the \VRI filters allowed to assess the
dust distribution in the coma of 67P. The gas contamination in the broad
band filters was negligible for all four epochs, as was verified in
spectra and measurements through an interference filter with the
passband centered in regions without any gas emission (central $\lambda
= 8340~\AA$, $\Delta \lambda = 480~ \AA$). No or negligible gas emission was present in this relatively wide passband, which was verified a posteriori also by the spectra.

\section{Data analysis}

\begin{figure*}
\centering
\includegraphics[width=9cm]{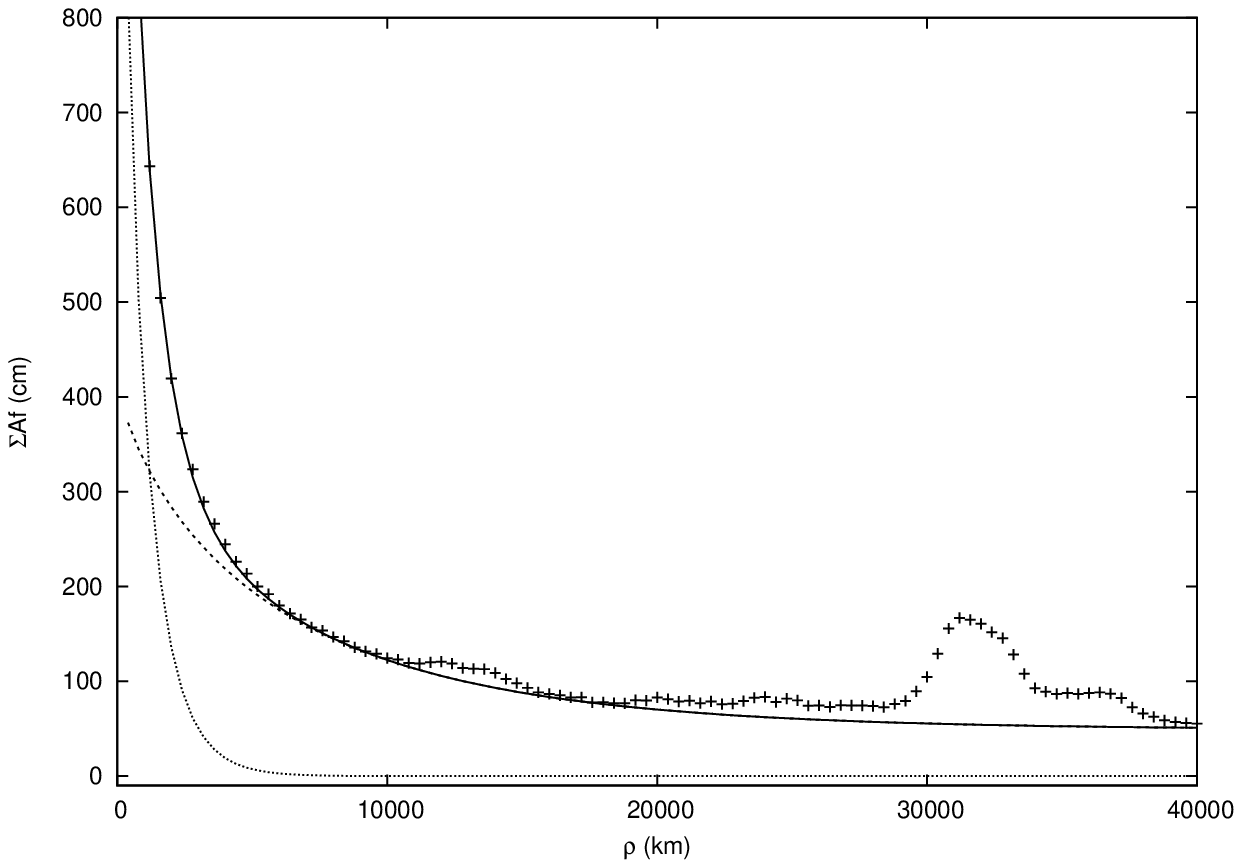}
\includegraphics[width=9cm]{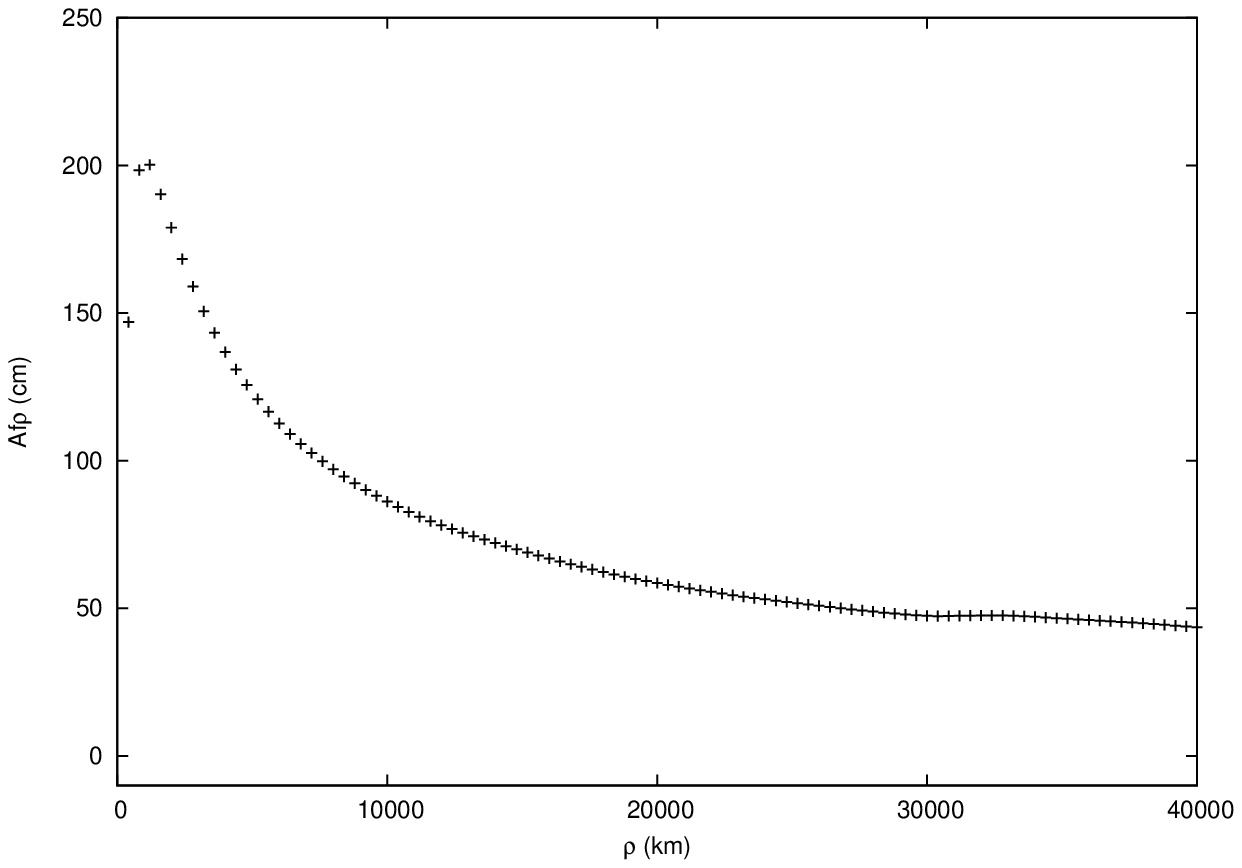}

\caption{
Left:  Measured \R\ filter \SAf\ in function of the projected nucleocentric distance $\rho$ for observations at 
\rh $\approx$ 2.3 AU together with the model functions. Crosses represent the measurements, the solid line shows the total model function, the dashed line gives 
the part related to \Nc+\Nl\ and the dotted line that of \Ns. The "bumps" (for instance, that at about 32000 km) are 
due to the remnant flux from background stars, easily identifiable by eye in the images. Note that the profile has been modeled up to $10^5$ km, but here a smaller part is shown. 
 Right: Measured \Afrho\ function for the same date and filter. This shows that the \Afrho\
is strongly dependent on $\rho$\ and consequently it cannot be used as proxy of the dust production}. 
\label{fig_SAf}
\end{figure*}

The final resulting images were initially analyzed using the \SAf($\rho$)
function. \SAf($\rho$) is proportional to the average column density
of the solid component at the projected nucleocentric distance
$\rho$. It is equal to $2\pi \rho Af(\rho)$ and is measured in cm. As
shown in \cite{Tozzi2007}, \SAf($\rho$) is constant with the projected nuclear distance $\rho$ in the case of a comet with a dust outflow of constant velocity and
production rate, and if sublimation or fragmentation of the grains are
excluded. Only the solar radiation pressure introduces a small linear
dependence, but usually only at large distances from the
nucleus, definitely larger than our field of view (FOV). Unfortunately, despite of all efforts, our final
images suffered from remnant flux contamination of the numerous
background stars so much that the stars were not completely erased
even after using the median average. However, it was always possible
to check and verify the radial profile of the \SAf~ function up to
about $10^5$ km from the nucleus. Figure \ref{fig_SAf}, left panel, gives an
example of a measured \SAf~ profile. Clearly, it is not constant with
radial distance $\rho$ from the nucleus and shows a fast decrease as $\rho$\ increases, until it reaches a constant value at about 20000 km. The \Afrho\ function, shown on the right panel of the same figure, is also strongly dependent on $\rho$ and clearly cannot be
used as a proxy of the dust production rate.

Using a trial \& error procedure the profiles were reproduced by
integrating along the line of sight of a spherical symmetric coma with
an ``optical density distribution'' of the dust as $ \Nd = [\Nc + \Nl e^{-(\frac{d}{\Ll})}+\Ns e^{-(\frac{d}{\Ls})}]/d^2$ where $d$
is the (non projected) nucleocentric distance in km. This function has a constant term
\Nc\ representing the constant optical density and two exponential terms  with \Ns\ and \Nl\ describing a
short-scalelength and a
long-scalelength optical density increases, respectively. The $d^2$ term
represents the spatial density attenuation due to geometric expansion
of the dust in the coma, \Ll~ and \Ls~ are the relative scalelengths
of the long and short-scalelength variations in \Nd. \Nc, \Nl,
\Ns\ describe the ``optical density distribution'', i.e. the albedo
multiplied by the light scattering area of the grains in a cubic
centimeter.  They are measured in $cm^{-1}$ (area over volume,
$cm^2/cm^3$). Of course, in order to derive the corresponding spatial
density of the dust (number of grains per cubic centimeter) one has to
consider the albedo and size distribution of the grains.  

For comparison with the data, \Nd\ is integrated along the line of sight. Figure \ref{fig_SAf} provides
an example of the computed profile as a function of the projected nucleocentric distance ($\rho$) using the adopted model function
\Nd.  The agreement between the measured radial profiles and the
models, defined above, is very good.  The only differences are within the "bumps" caused by the background stars.  The best
values of the model parameters \Nc, \Nl, \Ns, and \Ll~ and
\Ls\ together with the observing geometry and filters are listed
in Table \ref{tab_exp} for the four observing epochs. The errors indicated in the table refer mainly to 
the dominating systematic errors. \Nc\ is the most affected by the errors
due to contaminating flux from background stars. Our checks of the
\SAf\ profiles up to more than $10^5$ km, identifying the star
contribution visually, indicate an uncertainty of \Nc~ not greater
than a factor two. \Ns\ estimation is also affected by the systematic
errors because \Ls\ is of the order of the seeing, in particular for the observations taken during the first epoch. 
Hence, for that epoch, \Ls\ should just be considered as an upper limit and,
consequently, \Ns\ as a lower limit.  The parameters less affected by
the systematic errors are \Nl\ and \Ll, because \Ll\ is always much
larger than the seeing and the intensity of this component is high enough so that the
contribution of possible background stars is negligible.  By varying
these parameters in the adopted function for \Nd, we estimate that the
uncertainties are better or of the order of 25\% for \Nl~ and \Ll.


The last three columns of Table \ref{tab_exp}  
are the optical cross sections (SA), measured as the integral of \Nd,  of three components: constant(SA$_c$), long-scalelength (SA$_l$) and short-scalelength(SA$_s$). They represent 
the total area covered by the particles of each component multiplied by the particle albedo. At first order, SA is  
independent of the seeing and  gives "real" value (not just upper or lower limits) also for the short scalelength component. 
For the constant component, we arbitrarily computed its optical cross section assuming a limit for the nuclear distance of $5 \times 10^4$ km, since in principle it extends to infinity. The results show that, while SA of 
the constant component does not change much with \rh, the SA values of the other two components change 
a lot (by factors up to 20-40). We conclude that the two exponential components were replenished by
fresh grains produced as the comet approached the Sun.  However, we cannot exclude a small replenishment of 
fresh grains also in the constant component, because the grain ejection velocity should increase with 
decreasing \rh.

It is important to note that the flux contribution from the nucleus was not negligible when the comet was at \rh$\approx$ 3AU. 
With the radius $r_c$ = 2.0 km and the geometric albedo in the \R filter $A$ = 0.054 \citep{Lamy2006, Lamy2007, Kelley2009}, the optical
cross section (SA) of the nucleus at 0\degr\ phase angle is equal to 0.68 \km2.  Taking into account the factor 4 due to the different definition of albedo \citep{AHearn1984}, the optical cross section of the nucleus corresponds to 2.7 \km2, which is a large part of the measured SA$_s$ = 3.9 \km2 of the short-scalelength cross section in our \R filter. For the other epochs, the contribution of the nucleus is always negligible.

In the past, such anomalous enhancements were observed in other comets and interpreted by the presence 
of organic grains that sublimated in the coma with a certain lifetime $\tau$ while moving away from the nucleus 
\citep{Tozzi2004, Tozzi2007}. If this is the case the lifetime (and the scalelength) should decrease 
when the comet approaches the Sun. 
Our results for 67P show an increase of \Ll~ with the comet approaching the Sun, which excludes that grain sublimation or any other photolytic process were
responsible for the enhancement.

\begin{table*}
\caption{\label{tab_exp} Fit parameters of the \SAf~  profiles.  The first two columns give the average 
date of observations and the filter used (\V, \R, \I).
 \Nc, \Nl~ and \Ns~ are the optical density distributions of the constant, long and short
components, respectively. \Ll~ and \Ls~ are the respective equivalent scalelengths. The last three columns 
give the total optical cross sections of the three components. }
\begin{center}
\begin{tabular}{ccr@{$\pm$}lr@{$\pm$}lr@{$\pm$}lr@{$\pm$}lr@{$\pm$}lr@{$\pm$}lr@{$\pm$}lr@{$\pm$}l}
\hline
\hline
Date &Filter&\multicolumn{2}{c}\Nc               &\multicolumn{2}{c}\Nl                & \multicolumn{2}{c}\Ll&\multicolumn{2}{c}\Ns               &\multicolumn{2}{c}{\Ls}&\multicolumn{2}{c}{$SA_c$}&\multicolumn{2}{c}{$SA_l$} & \multicolumn{2}{c}{$SA_s$}\\
YYMM &      &\multicolumn{2}{c}{$10^{-8}cm^{-1}$}&\multicolumn{2}{c}{$10^{-8}cm^{-1}$}&\multicolumn{2}{c}{$km$}&\multicolumn{2}{c}{$10^{-8}cm^{-1}$}&\multicolumn{2}{c}{$km$}&\multicolumn{2}{c}{$km^2$}&\multicolumn{2}{c}{$km^2$}    &\multicolumn{2}{c}{$km^2$}\\
\hline
0806&  \V  &  0.02 & ${^{0.02}_{0.01}}$  &   0.30 & 0.07   &   2930 & 60   &    0.42 & 0.10     &   793 & 16  &  14 & $^{14}_{7}$   &  8.0 & 2.0  &  3.0 & 0.6 \\
0806&  \R  &  0.02 & ${^{0.02}_{0.01}}$  &   0.19 & 0.05   &   3750 & 75   &    0.40 & 0.10     &  1056 & 21  &  14 & $^{14}_{7}$   &  6.5 & 1.6  &  3.9 & 1.0 \\
0806&  \I  &  0.02 & ${^{0.02}_{0.01}}$  &   0.31 & 0.08   &   3170 & 63   &    0.25 & 0.06     &   800 & 16  &  15 & $^{14}_{7}$   &  9.0 & 2.2  &  1.7 & 0.4 \\
0809&  \V  &  0.04 & ${^{0.04}_{0.02}}$  &   0.31 & 0.08   &   7350 & 145  &    0.99 & 0.25     &   976 & 20  &  30 & $^{30}_{15}$  &  20  & 5    &  8.9 & 2.2 \\
0809&  \R  &  0.03 & ${^{0.03}_{0.015}}$ &   0.26 & 0.06   &   9300 & 186  &    0.74 & 0.18     &  1330 & 27  &  24 & $^{24}_{12}$  &  23  & 6    &  9.3 & 1.8 \\
0809&  \I  &  0.03 & ${^{0.03}_{0.015}}$ &   0.26 & 0.06   &   9600 & 190  &    0.77 & 0.15     &  1390 & 28  &  23 & $^{23}_{12}$  &  23  & 6    &  10.3& 2.6 \\
0810&  \V  &  0.06 & ${^{0.06}_{0.03}}$  &   0.26 & 0.06   &  13000 & 260  &    2.08 & 0.52     &  1300 & 26  &  44 & $^{44}_{22}$  &  32  & 8    &  26  & 6 \\
0810&  \R  &  0.13 & ${^{0.13}_{0.06}}$  &   0.53 & 0.13   &  13000 & 260  &    3.43 & 0.61     &  1380 & 28  &  92 & $^{92}_{46}$  &  66  & 16   &  45  & 11 \\
0810&  \I  &  0.11 & ${^{0.11}_{0.05}}$  &   0.50 & 0.12   &  11300 & 226  &    2.82 & 0.70     &  1380 & 28  &  79 & $^{79}_{38}$  &  54  & 13   &  37  & 9 \\
0901&  \V  &  0.05 & ${^{0.05}_{0.02}}$  &   0.42 & 0.10   &  62000 & 1240 &    2.38 & 0.60     &  4300 & 86  &  40 & $^{40}_{20}$  &  282 & 70   &  96  & 24 \\
0903&  \R  &  0.03 & ${^{0.03}_{0.015}}$ &   0.41 & 0.10   &  46600 & 930  &    2.51 & 0.63     &  4700 & 94  &  28 & $^{28}_{14}$  &  206 & 51   &  111 & 28 \\
\hline
\end{tabular}
\end{center}
\end{table*}

\begin{figure}[]
\centering
\includegraphics[width=9cm]{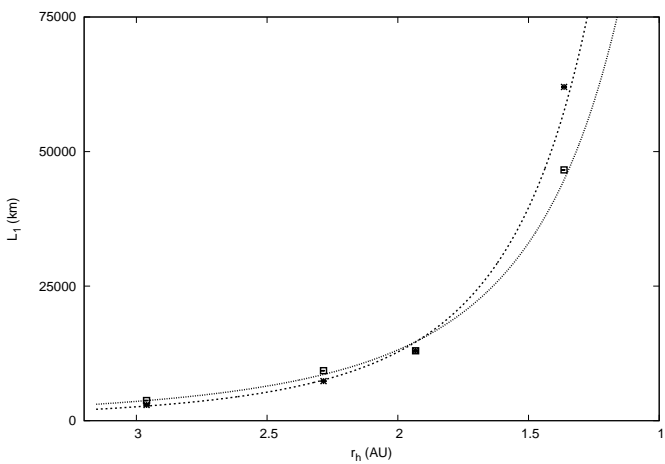}
\caption{\label{fig_L_rh}
Measured \Ll~ for \V (star) and \R (square) filters as a function of the heliocentric distance.
The lines show the best fit assuming the function described in the text. Solid line
represents the \V and  dotted line the \R filter.} 
\end{figure}

Figure \ref{fig_L_rh} gives the scalelength \Ll~ as a function of the heliocentric distance \rh. The scalelength values  are well fitted by the function function \Ll = $c r_h^{\gamma_1}$ with  best fit given for 
c =  $(195000\pm29000)$  km and $\gamma_1 = (-3.93\pm0.20)$ in \V
and 
c = $(120000\pm18000)$ km and  $\gamma_1 = (-3.20\pm0.19)$ in \R.

\begin{figure}[]
\centering
\includegraphics[width=9cm]{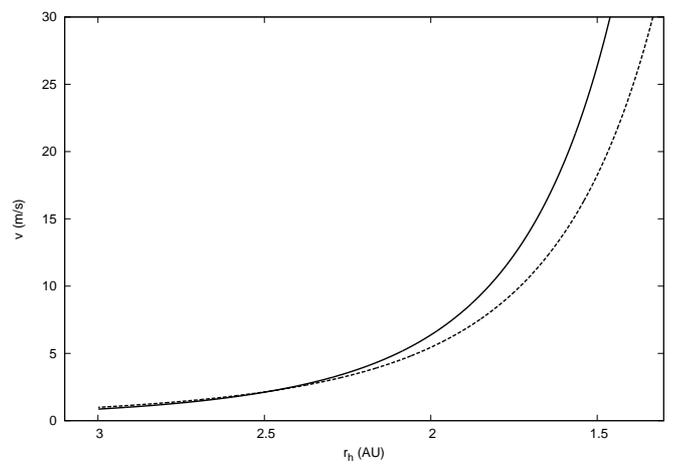}
\caption{\label{fig_V_rh}
Computed grain ejection velocity as function of the heliocentric distance for observations in \V (solid line) and in \R (dotted line).} 
\end{figure}

If we assume that this long-scalelength component was due to fresh dust produced in a 
spherically symmetric coma during the approach of the comet to the Sun, the coma was expanding  with a velocity given 
by $ \frac{d \Ll}{dt}$, which is equal to $ c \times \gamma_1 \times (r_h(t))^{(\gamma_1 -1)} \times \frac{d r_h(t)}{dt}$.
Since in this part of the orbit the heliocentric distance varies linearly with time as  $\rh = r_{h0} + v_0 
T_p$ with $T_p$ = days to perihelion, $r_{h0} = 1.011$ AU and $v_0 = -7.18 \times 10^{-3}$ AU/day,  the time variation of the scalelength is $ \frac{d L_1(t)}{dt} = c \times v_0 \times \gamma_1 \times (r_h(t))^{(\gamma_1 -1)} $.
This is the expansion velocity of the grains that are at nucleocentric distance equal 
to \Ll. It is known that the grains are accelerated by gas drag only in the inner part of the coma, i.e. within few radii of the comet 
nucleus (see \eg \cite{Combi1997}) and that they move with almost constant velocity beyond that limit. They  are affected only by solar radiation pressure, efficient at large scales, and by the gravity of the comet, efficient only 
very close to the nucleus. This means that, roughly speaking, the grains expand in the coma 
with a widely distributed, but unchanging velocity. Hence, the grains at nucleocentric distance \Ll\ were ejected some time before ($\Delta T$) with the velocity  $ \frac{d L_1(t)}{dt}$. To compute $\Delta T$ we have just to divide \Ll\ by $ \frac{d L_1(t)}{dt}$. This gives $ \Delta T = \frac{1}{v_0 \times \gamma_1} \times \rh(t) $, \ie the time necessary to reach \Ll\ is linearly dependent on \rh. For example, for observations in the \R filter, $\Delta T$ is about 87 days at \rh = 2 AU. The  value of $ \frac{d L_1(t)}{dt}$ for the same \rh\ and filter is about 150 km/day or 1.73 m/s, i.e. the expansion velocity is quite low. As the grains are expanding almost freely (see above), the expansion velocity at a certain time $T_1$ is equal to the ejection velocity at time $T_1 - \Delta T$. By solving the equations numerically, we get the ejection velocities as  functions of the heliocentric distance, as shown in Figure \ref{fig_V_rh}. They are equal to $ V_{ej} = V_{ej0}\times  r_h^{\gamma_1-1} = V_{ej0}\times  r_h^{\gamma}$, with $ V_{ej0} = 195\pm21$ and $100\pm11$ m/s in \V\ and \R, respectively.

This formula is similar to the ejection velocity assumed in models, \ie  $ V_{ej} \propto \sqrt{\beta} \times  r_h^{\gamma} $, with $\beta$ = the ratio of the radiation pressure over solar gravity forces (see \eg \cite{Fulle2010}). We get $\gamma$ = -(4.93$\pm$0.20) and -(4.20$\pm$0.20) in \V and \R, respectively, which is quite different from the assumed values: -0.5 \citep{Ishiguro2008, Kelley2008, Kelley2009} or -3 \citep{Agarwal2010}. The ratio of $\beta$, of the particles observed in \V with respect to those observed in \R is about 3.8$\pm$1.2, a rather high value that is difficult to explain.  

It is important to note that all results about the grain ejection velocity assume isotropic outflow of the dust. If the comet showed an  asymmetric emission (e.g., a jet) in the direction of the observer, this would have changed the results leading to a higher ejection velocity. However, this would mean that the jet direction  and intensity had to remain unchanged for about four days during each epochs, since even the images taken several days apart during each epoch did not show any change in flux. Such a stable jet could only be produced if the spin axis of the nucleus was pointing exactly into the observer's direction, which could be the case only for one of the three epochs if at all, but certainly not for all three epochs.

\section{Discussion}

\subsection{Other Observations} 

It is important to note that during the post-perihelion phase comet 67P displayed a completely different 
behavior from what we found in the pre-perihelion images. The \SAf~ profiles did not show any anomalous 
enhancement towards the nucleus. For example, the profiles obtained from HST observations (see \cite{Lamy2006}) and downloaded from the HST archive, show only the signature of the nucleus itself at $\rho$ close to zero, but are otherwise 'flat' up to $\rho = 4\times 10^4$ km, the FOV of the images. A similar behavior is also found in other post-perihelion measurements, which demonstrates that the slope parameters of log(I) vs log($\rho$) are equal to about -1 \citep{Lara2005, Schleicher2006}. Obviously, after perihelion the comet had already developed an extended coma with the conditions necessary for a reliable determination of \Afrho\ (see above). 

\subsection{Grain size}

As shown above, for heliocentric distances greater than $\approx$ 2 AU, the dust expansion velocities in the coma of 67P were about two orders of magnitude smaller than those usually claimed for cometary comae (see \cite{Crifo1997}, \cite{Foster2007} and references therein). This may be an indication that the grains in 67P were big. \cite{Agarwal2007, Agarwal2010} computed that velocities of the order of 30-50 m/s at \rh = 1.3 AU correspond to grain radii ranging from about 10 $\mu$m to 1 mm (depending on the emission scenario) assuming spherical grains with a density equal to 1 g/cm$^3$. Since it is likely that the grains are fluffy and non spherical (and the density much lower than than assumed), their size can be much larger. \cite{Hadamcik2010} have also reported an anomalous slope log(I) vs log($\rho$) (close to -1.5) in the inner coma ($\rho$ between  2000 and 8500 km) in December 2008, which they interpreted as being due to the presence of large particles. The presence of large and fluffy particles is also consistent with polarization measurements performed in March 2009.  \cite{Hadamcik2010} note that the high polarization (~6\%) observed near the nucleus is typical for small, sub-micron particles (cf. comet Hale-Bopp in \cite{Hadamcik2003}). This contradiction can be resolved by suggesting large, but porous particles \citep{Hadamcik2010}. Note that several authors \citep{Fulle2004, Fulle2010, Moreno2004, Kelley2008, Ishiguro2008}) suggested the presence of grains in the cm-range based on the observations of a trail, the tail and necklines in 67P.

\subsection{Grains Density in the Coma}

For regions with $d \ll \Ls \ll \Ll $, which are relevant for the {\it Rosetta} spacecraft during its close approach or when  orbiting  the nucleus of 67P, the mean optical density distribution of the dust at a nucleocentric distance $d$ is simply equal to $\Nd = (\Nc + \Nl+ \Ns)/d^2 = \N0/d^2$, with \N0\ represents the optical density distribution at $d$ = 1 km. For example, at 1.93 AU from the Sun, we find \N0 = $4.1 \times 10^{-8} cm^{-1}$ (see the \R filter results in Table \ref{tab_exp}).
Assuming the power-law size distribution $ n(r) = n_0 (\frac{r}{r_0})^{-k}$, equal for the three components of \Nd, the optical density  is equal to
$\Nd  = \int{ \frac{\pi r^2 A n_0 }{d^2} \frac{r^{-k}}{r_{0}^{-k}} dr} = \frac{\pi A n_0 }{d^2 r_{0}^{-k}} \int{ r^{2-k} dr}  = \frac{\pi A n_0 }{d^2 r_{0}^{-k} (3-k)} [{r_{x}^{3-k}-{r_{m}^{3-k}}}]$
with $A$ = the grain albedo, and the integral extended from the minimum ($r_m$) to the maximum ($r_x$) radius of the grains assumed to be spherical.
The local density mass due to the dust is 
$M(d) = \int{ \frac{4 \pi r^3 \sigma n_0 }{3 d^2} \frac{r^{-k}}{r_0^{-k}} dr} = \frac{4 \pi \sigma n_0 }{3 d^2 r_0^{-k}} \int{ r^{3-k} dr}  = \frac{4 \pi \sigma n_0 }{3 d^2 r_0^{-k} (4-k)} [{r_x^{4-k}-{r_m^{4-k}}}]$
with $\sigma$ being the average density of the grains.
The ratio $\frac{M(d)}{\Nd}$ is then independent on $d$ and is equal to $ \frac{4}{3} \frac{\sigma}{A} \frac{3-k}{4-k} \frac{[{r_x^{4-k}-{r_m^{4-k}}}]}{[{r_{x}^{3-k}-{r_{m}^{3-k}}}]}$.
Assuming $r_x \gg r_m$ and $ k < 3$, as big grains dominate, we have $M(d) = \Nd \frac{4}{3} \frac{\sigma}{A} \frac{3-k}{4-k} r_x$. 
So, for the given physical parameters of the grains (albedo and density), the ratio $\frac{M(d)}{\Nd}$ depends on the size of the biggest grains in the dust size distribution, $r_x$ and, weakly, on k. With a grain density  $\sigma = 0.2~ g/cm^3$, an albedo $A = 0.04$, and $k =2$ the mass density is $ M(d) = 3.3 \Nd r_x$. 
For example, for \rh $\approx$ 2 AU, at  10 km from the nucleus, the optical density is 
$4.1 \times 10^{-10} cm^{-1}$. Assuming a maximum grain size of 1 mm, the mass density would be 
$1.4 \times 10^{-10} g/cm^3$. Hence, during one orbit revolution at this distance each square meter of the {\it Rosetta} spacecraft 
would intercept a total mass of only 8.6 g. If the maximum grain radius is 5 cm the mass density would be $6.5 \times 10^{-9} g/cm^3$ and the mass intercepted would be 430 g, a non-negligible amount, considering that the maximum 
cross section of the spacecraft is 17 $m^2$.

It is important to note that the main contribution to the density of the coma in regions close to the nucleus is produced by the short scalelength component, that sometines is affected by the
seeing of the observations. In this case N$_s$ may be a lower limit. However, for the considered example at
\rh $\approx$ 2 AU this component is not strongly affected, because its scalelength corresponds to 
about 1.3\arcsec, which is larger than the seeing of the observational nights.

\section{Conclusions} 

The anomalous enhancement of the dust density of the coma of 67P towards the nucleus, measured in
2008 and 2009, can be explained by a very slowly expanding dust cloud, with velocities of the order of 1 m/s 
at \rh=3 AU. The slow dust expansion velocity supports a scenario where at large pre-perihelion heliocentric distance the coma of 67P was dominated by large grains, with dimensions greater than 10 $\mu$m - 1 mm, if they are spherical with a density equal to 1 g/cm$^3$, or much larger, if they are fluffy aggregrates.
By modeling the radial coma flux profiles with a spherically symmetric coma, the optical density distribution of 
the dust was quantified as a function of the nucleocentric distance $d$. The ejection velocity was derived 
as a function of the heliocentric distance, \rh. The model also allowed to derive the mass density  of the dust as a function of $d$ assuming 
 a grain size distribution in which big grains dominate the coma. 

{}


\end{document}